# Semi-Automated Quality Assurance in Digital Pathology

Tile Classification Approach

Meredith VandeHaar, SCT(ASCP), Clinch, M., Yilmaz, I., Rahman, M.A., Xiao, Y., Dogany, F., Alazab, H.M., Nassar, A., Akkus, Z., Dangott, B.

# Abstract

Quality assurance is a critical but underexplored area in digital pathology, where even minor artifacts can have significant downstream effects. Artifacts such as focus distortion, digital compression, fingerprints, squamous cells from slide handling, and dust analogs have been shown to negatively impact the performance of AI diagnostic models. In current practice, trained staff manually review digitized images prior to release of these slides to pathologists who are ultimately responsible for final assessment of the quality of the slides they use to render a diagnosis. Conventional image processing approaches, such as those implemented in HistoQC (Janowczyk et al., 2019), provide a foundation for detecting artifacts on digital pathology slides. However, tools like HistoQC do not leverage deep learning, which has the potential to improve detection accuracy and scalability. Despite these advancements, comprehensive methods for artifact detection and quality assurance in digital pathology remain limited, presenting a critical gap for innovation.

We propose an AI algorithm designed to screen and classify digital pathology slides by analyzing each tile and categorizing it into one of 10 predefined artifact types or as background (negative). This algorithm identifies and localizes artifacts across the slide, creating a detailed map that highlights regions of interest. By directing human operators to specific tiles affected by artifacts, the algorithm minimizes the time and effort required to manually review entire slides for quality issues.

From internal archives and The Cancer Genome Atlas, 133 whole slide images were selected and 10 artifacts were annotated using an internally developed software ZAPP (Mayo Clinic, Jacksonville, FL). Ablation study of multiple models at different tile sizes and magnification was performed. InceptionResNet was selected. Following several model trainings, performance for each artifact varied greatly. Single artifact models were trained and tested, followed by a limited multiple instance model with artifacts that performed well together (chatter, fold, and pen). From the results of this study we suggest a hybrid design for artifact screening composed of both single artifact binary models as well as multiple instance models to optimize detection of each artifact.

Implementing an AI driven artifact screening method has the potential to decrease the amount of time required of humans to detect quality issues by forwarding only the WSIs likely to contain them for manual review.

# Semi-Automated Quality Assurance in Digital Pathology

Tile Classification Approach

Meredith VandeHaar, SCT(ASCP), Clinch, M., Rahman, M.A., Xiao, Y., Yilmaz, I., Nassar, A., Akkus, Z., Dangott, B.

# Introduction

Pathology, as a discipline, has deep historical roots stretching back thousands of years. Even in ancient Egypt, practitioners documented pathological changes on papyrus, most famously in the context of embalming procedures that served both religious and medical purposes. Early records indicate observations of organ abnormalities, but without much evidence to suggest systematic understanding of disease processes. Centuries later, in classical Greece, Hippocrates introduced the humoral theory, which systematically approached pathology as an imbalance of bodily fluids—a view that dominated Western medicine for centuries.

The true emergence of modern pathology, however, can be traced to the work of Rudolf Virchow and his contemporaries in the 19th century. Often regarded as the father of cellular pathology, Virchow established the principle that diseases arise primarily from changes within cells, laying the foundation for microscopic pathological practice. Advances in tissue processing soon followed. In 1869, the introduction of paraffin embedding improved the structural preservation of tissue samples, and in 1893, formalin fixation further enhanced the clarity of cellular and tissue morphology. These innovations enabled pathologists to perform detailed, systematic descriptions of diseases at the cellular level.

Throughout the 20th century, pathology continued to refine techniques, such as immunohistochemical staining and in situ hybridization, which allowed for more precise identification of molecular markers within tissue sections. Yet one crucial element remained essentially unchanged for decades: the fundamental practice of visually inspecting thin slices of formalin-fixed, paraffin-embedded (FFPE) tissue stained with hematoxylin and eosin (H&E) on a glass slide under a light microscope. This time-honored approach is still the bedrock of diagnostic pathology, despite the many technological advances that have otherwise revolutionized medicine.

Over the past decade, however, digital pathology has ushered in a major shift, driven in large part by improvements in whole slide imaging and the rapid growth of artificial intelligence applications. In 2017, the U.S. Food and Drug Administration granted its first approval for primary diagnosis using a digital pathology system, the Philips IntelliSite Pathology Solution, marking a turning point for fully digitized workflows (Philips, 2017). Other platforms have since followed, and many pathology departments worldwide are investing in digital capabilities.

Nevertheless, not all digital pathology data is created equal. Faced with an ever-increasing number of slides, laboratories often struggle to ensure that scanned images meet rigorous quality standards. Poor-quality slides—due to staining issues, physical artifacts, or digitization errors—frequently make their way into institutions' digital tissue registries. Over time, these subpar slides can impede research and affect diagnostic consistency and the performance of digital diagnostic aids, such as those driven by AI (Schömig-Markiefka et al., 2021), underscoring the need for robust quality control measures. Even as the field embraces digital innovation, the history of pathology demonstrates that some core practices have remained remarkably consistent over the past century, emphasizing the importance of careful integration of new technology into longstanding traditions of morphological assessment (van den Tweel & Taylor, 2010).

## Digital Pathology Workflow: From Collection to Digital Storage

The pathology process starts with the patient in a procedure room or surgery when tissue is taken from the patient. Small needle, punch, or resection biopsies formalin. Larger specimens, such as whole organ resections, are transported to the lab to be dissected and representative parts selected for subsequent fixation in formalin. Tissue sections and small biopsies are placed in plastic cassettes and onto tissue processors where they are dehydrated over a few hours when the water is slowly replaced with paraffin wax. After this process is complete, these tissue samples are physically oriented and embedded in paraffin wax along with the cassette bearing specimen and patient identification information. Next, histologists cool the paraffin block with ice and thinly slice the tissue on a microtome into delicate slices as thin as 4μm. These are floated on a water bath, and a properly identified slide is submerged and raised to meet one or more of them from below. Slides are then run through an oven to melt the paraffin wax and adhere the tissue to the slide. Slides with adhered tissue are then run through various staining solutions to add differential color to various components of the tissue. Finally, after adding mounting media and a coverslip, these prepared slides can be digitally scanned and stored on a data server either on premise, or in the cloud.

Quality assurance is a critical but underexplored area in digital pathology, where even minor artifacts can have significant downstream effects. Artifacts such as focus distortion, digital compression, fingerprints, squamous cells from slide handling, and dust analogs have been shown to negatively impact the performance of AI diagnostic models. Schömig-Markiefka et al. (2021) demonstrated that the effect of these artifacts increases with image compression levels, severity of focus aberration, and the severity and number of artifacts in the case of physical contamination underscoring the importance of rigorous quality control to maintain AI model accuracy.

Current uses of deep learning to detect rare events such as mitoses, and accurately classify malignancies require a large amount of high quality supervised datasets. These are available through various challenges (TIGER [https://tiger.grand-challenge.org/], MITOS-ATYPIA-14 [https://mitos-atypia-14.grand-challenge.org/] (Chen et al.,2016), and others) are readily available. However, high quality datasets of digital pathology defects are not, leading researchers to curate their own. Most studies have been conducted with small, noisy, and weakly supervised datasets including slides processed in a manner to artificially increase the likelihood of artifacts (Cui et al., 2021) (Foucart et al., 2016). Still other researchers have resorted to augmenting datasets with artificially generated artifacts (Jurgas et al., 2023) (Schömig-Markiefka et al., 2021).

In current practice, trained staff manually review digitized images prior to release of these slides to pathologists who are ultimately responsible for final assessment of the quality of the slides they use to render a diagnosis. Large artifacts encompassing a significant portion of the slide are trivial to detect, however, smaller artifacts may not be found until the time of diagnosis by a pathologist. If a slide is deemed unsuitable, it may be sent back to the laboratory for re-preparation, which can add a full day to the case's turnaround time. Alternatively, pathologists may choose to make a diagnosis from another slide or region of better quality where possible, leaving additional slides or regions of poor quality in the digital tissue registry. These neglected slides may later be selected for downstream analysis, potentially propagating the impact of quality issues. (Browning et al., 2024)

Conventional image processing approaches, such as those implemented in HistoQC (Janowczyk et al., 2019), provide a foundation for detecting artifacts on digital pathology slides.

However, tools like HistoQC do not leverage deep learning, which has the potential to improve detection accuracy and scalability. Despite these advancements, comprehensive methods for artifact detection and quality assurance in digital pathology remain limited, presenting a critical gap for innovation.

## Quality Assurance in Digital Pathology

The quality of each step in the preparation and processing workflow significantly influences the ultimate quality of digital pathology images. A multitude of potential errors can arise, impacting both the immediate imaging results and downstream computational analyses. During the collection, grossing, and fixation stages, issues such as air-drying fresh specimens prior to formalin fixation, excessive cold ischemic time, suboptimal fixation duration, variability in tissue processor settings or reagents, and physical damage to delicate sections can degrade slide quality. Additional complications may occur during decalcification, where excessive or insufficient treatment of bone-containing sections impacts tissue integrity, or through contamination from microscopic tissue remnants from prior cases.

In the embedding process, improper tissue orientation and contamination from tools can further compromise results. During microtomy, factors such as incorrect paraffin block temperature, dull blades, incorrect section thickness, and contamination from water baths can cause tears, folds, or distorted sections. Staining, mounting, and coverslipping introduce further risks, including environmental debris, misalignment, excess mounting media, variation in staining reagents and timing, or damage to polymer coverslips by mechanical abrasion.

Retrospective or ancillary slides face additional challenges, such as the accumulation of dust, fingerprints, and scratches from slide handling or storage. In the scanning and data processing phase, suboptimal focus, undetected tissue regions, misaligned tile stitching, improper color calibration, or image compression can reduce image clarity. Mechanical factors such as lens contamination, vibration, barcode failure, and improper slide handling can exacerbate these issues, emphasizing the critical need for meticulous quality control at every step of the digital pathology workflow.

## Proposed AI-Driven Artifact Detection

We propose an AI algorithm designed to screen and classify digital pathology slides by analyzing each tile and categorizing it into one of 10 predefined artifact types or as background (negative). This algorithm identifies and localizes artifacts across the slide, creating a detailed map that highlights regions of interest. By directing human operators to specific tiles affected by artifacts, the algorithm minimizes the time and effort required to manually review entire slides for quality issues.

The goal of this targeted approach is to streamline the quality control process, enabling operators to focus on areas that need attention, such as regions with folds, tears, or debris, and increase likelihood of detection of artifacts that are more difficult for a human operator to detect, such as areas of poor focus. By doing so, the algorithm not only accelerates the review process, but also reduces the likelihood of human error caused by fatigue or oversight. Furthermore, it ensures consistent and objective artifact classification across slides, facilitating standardized quality assessments. The first step in this pipeline is screening with a faster tile classification algorithm followed by a segmentation algorithm to precisely identify artifact locations on positively screened slides. This paper focuses on the former algorithm.

# Methods

## Dataset

We initially collected 79 digital pathology slides from internal archives at two sites, scanned using multiple scanner models. Additionally, 59 slides were sourced from The Cancer Genome Atlas (TCGA) to enhance data diversity. The images were split on a slide level into 60:20:20 proportions for training, validation, and testing, ensuring no overlap between splits. Eleven artifact classes (air bubble, chatter, coverslip scratch, debris, dropped tissue, dust, focus, fold, pen, tissue scratch, and coverslip edge) were annotated by a panel of five reviewers, consisting of a cytotechnologist, a pathologist, a data scientist, and two data science trainees with expertise in pathology data. Coverslip edge was excluded because only one slide was available with an example of the artifact. One Papanicolaou stained cytology smear, and three immunohistochemically stained slides were also excluded leaving 133 H&E stained slides.

## Data Annotation

Data were annotated using ZAPP (Mayo Clinic, Jacksonville, FL), a custom Flask web tool. Using mouse or tablet input, polygonal annotations are stored as sets of coordinates in JSON format. Each annotation class was stored as an integer. WSI were then divided into several predesignated tile sizes at different magnifications for training the models in Python (3.7) using deepzoom functions from OpenSlide Python (1.1.2). Images and masks were stored within TFRecord files, and the class with the most area other than background was stored as the overall label of the image tile if it met the designated threshold area percentage for the class.

# Model Selection & Pipeline Strategy

To identify the optimal model architecture, tile size, and zoom level, an initial ablation study was conducted using an exhaustive grid search methodology. This study varied tile size (128 px², 256 px², 512 px²), zoom level (L0, L2, L4), and model architecture (VGG16 (Simonyan & Zisserman, 2016), InceptionV3 (Szegedy et al., 2015), DenseNet (Huang et al., 2017), ResNet (He et al., 2015), and InceptionResNet (Szegedy et al., 2016). Among the models, InceptionV3, InceptionResNet, ResNet50, and VGG19 showed roughly equivalent performance. InceptionResNet was selected due to its hybrid design combining the benefits of the Inception module's multi-scale feature extraction and ResNet's skip connections, which address the vanishing gradient problem and enhance learning for deeper architectures.

The optimal combination for the InceptionResNet model was determined to be a tile size of 256 px² at zoom level L2. This configuration effectively balanced image resolution and computational efficiency, capturing sufficient contextual detail at the chosen tile size and zoom level.

Following model selection, additional exhaustive grid search experiments were conducted to further investigate the feasibility of detecting individual artifact classes and to refine the pipeline for an ensemble model approach. Binary classification models were trained for each artifact class (*air bubble*, *chatter*, *coverslip scratch*, *debris*, *dropped tissue*, *dust*, *focus*, *fold*, *pen*, and *tissue scratch*) against the background class.

These experiments systematically varied tile sizes (128 px², 256 px², 512 px²) and magnification levels (L2, L4) to address two objectives:

1. **Feasibility Testing:** Determine whether artifacts with low performance in the initial ablation study, such as *focus*, *dust*, and *dropped tissue*, could be more effectively detected with alternative configurations.

2. **Optimization for Ensemble Modeling:** Identify the optimal tile size and magnification level for each artifact class to inform the design of an ensemble model pipeline, where specific configurations could be tailored to improve detection of challenging artifacts.

By isolating artifact classes in this manner, the experiments aimed to maximize individual artifact detection performance while exploring the potential for a modular and adaptive approach.

## Model Optimization

Further adjustments were made to optimize model performance in the multiple instance model. The annotation strategy was revised for *air bubble* and *dropped tissue* artifacts to improve their detection. Additional annotations were added for regions commonly misclassified as false negatives, such as distinguishing *tissue ink* from *pen* (Fig. 4 A & B). Other annotations were shrunk or refined to minimize excess background inclusion and limit imbalance of this class.

Following these adjustments, a new multiple instance model was trained using the InceptionResNet architecture with L2 magnification and tile size 256 px² (Figure 3). The top-performing artifact-specific models were then integrated into a smaller, refined multiple instance model, maintaining the same parameters (Figure 6), to streamline performance and efficiency for detecting artifacts across classes. Within the training set, for each artifact class, tiles from slides that were more than a factor below the mean were copied to meet, but not exceed the mean for that class to ensure that minority representations of the artifacts are learned. This step improved the performance for pen, fold, and chatter artifacts, but decreased performance for most others (Figure 5).

# Results

## Model Selection

InceptionV3, InceptionResNet, ResNet50, and VGG19 showed roughly equivalent performance ahead of Xception, DenseNet, and VGG16 (Table 2). InceptionResNet was chosen due to its features addressing the vanishing gradient problem, complex feature extraction. We were confident that it is better equipped to learn the features in our data as the dataset increases in size and complexity.

## Single Artifact Models

Artifact models trained on L2 achieved true positive rate above 90% for *air bubble, chatter, coverslip scratch, debris, dust, focus,* and *pen* on all three tile sizes 128, 256, and 512 px². Models for *dropped tissue* achieved a true positive rate of 90% or better for 256 and 512 px², and 31.5% true positive rate on 128 px². The models for *fold* and *tissue scratch* showed a similar pattern achieving 64.1% and 55.3% respectively on 128 px².

Models trained on L4 tiles showed poorer performance with increasing tile size. Models for *air bubble*, *dust, fold, pen,* and *tissue scratch* achieved true positive rate below 90% for detecting artifact (false negatives), background (false positives), or both at tile sizes 256 and 512 px² (average 87.8%, 73.8% respectively). This is likely due to the decreasing tile count with increasing tile size (average count 46.1, 28.3, and 20.4 for 128, 256, and 512 px² respectively)

## Limited Multiple Instance Model

To explore the feasibility of a hybrid multiple instance classification approach to tile classification screening, a second multiple instance model was trained with artifacts with significant morphologic, chromatic, and size distinction (pen, chatter, fold; Figure 6). This model shows an excellent true positive rate for pen and fold.

# Discussion

Implementing an AI driven artifact screening method has the potential to decrease the amount of time required of humans to detect quality issues by forwarding only the WSIs likely to contain them for manual review. A segmentation step following this screen could also be implemented to more precisely locate the artifact regions. This may allow for storage of a separate mask file that may be loaded to augment further downstream AI analysis by allowing them to be excluded. To date, our study is the largest study in terms of the number of artifact slides total. We have also included both internal slides from two scanners (Leica AT2, Grundium Ocus) and public data from an unknown mix of scanners in this study.

The nature of this, and other digital pathology studies involving multiple instance learning on WSIs is that a tile may contain more than one annotated class, but the model requires training data where tiles and labels are one-to-one in order to be scored perfectly. However, upon examination of requisite data, a classification error may still be considered accurate by a human in-the-loop. Therefore, the task of identifying artifacts accurately is not as important as distinguishing artifacts versus background tiles in the context of a human reviewer. Identifying which artifacts are present at the slide level to a sufficient percentage of total slide tiles may be useful in triaging tiles to be analyzed by a subsequent single artifact segmentation algorithm. The results of subsequent algorithms may guide a user in deciding whether to reprocess, and what steps in pathology processing should be reperformed (reembed, recut, restain, recoverslip, etc.) Because of this, we examined the results when scoring a true positive when any artifact is detected for a tile whose ground truth is any other artifact (Table 4). Adjusting scoring in this manner boosted the accuracy for nearly all artifacts.

Air bubbles may also be challenging to detect because the majority of them occur in whitespace on H&E stained slides. They are defined by a hard edge which may be easy to detect, but on tiles away from that hard edge, it is difficult even by the human eye to detect whether that tile is affected by an air bubble or not without a broader context. Air bubbles over tissue, however, cause a visual disruption colloquially termed "cornflaking" due to the characteristic brown color and rough texture imparted upon affected cells. This may be easier to detect, but examples in the dataset are few in number and difficult to locate as often slides affected by this issue are re-coverslipped. It is likely the best approach to procure more data to train a high context model (L4-1024). Because of the nature of such large artifacts requiring high context for a deep learning model to recognize consistently (air bubble, dropped tissue), future work on this project should include additional investigation and data collection for slides containing air bubbles as the number of slides needed to train the model sufficiently will require a lower magnification to gain the higher context. Lowering the magnification improves training time at the trade-off of sufficient data to train each class (Figure 7). Cui et al. also found that adding lower resolution data produced more false positive predictions as at lower resolution, most training tiles tend to have artifacts on them.

Using larger tiles worked better for deeper networks, but had a negative effect on shallower networks. A tile screening method utilizing a hybrid ensemble and multiple instance learning method is likely the most reasonable approach.

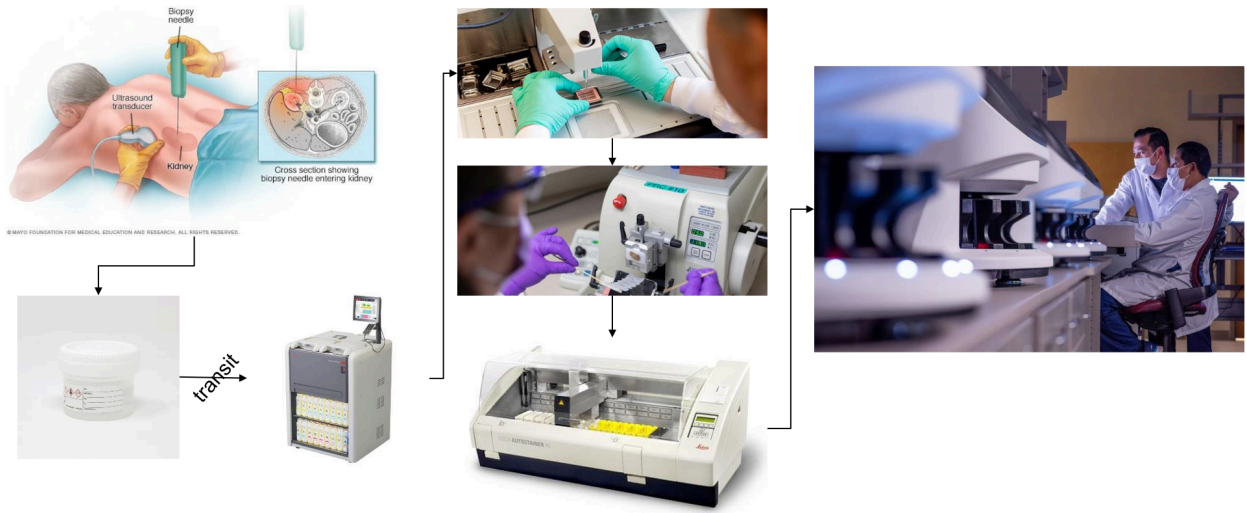

**Figure 1** illustrates the end-to-end workflow in digital pathology: (A) specimen collection, (B) transport/fixation, (C) tissue processing, (D) tissue embedding, (E) microtomy, (F) staining, mounting, coverslipping, and (G) digital slide scanning. The quality of each step dictates the quality of the scanned image. (Images: A, D, E, G, Mayo Clinic; B, Fisher Scientific; C, F, Leica Biosystems)

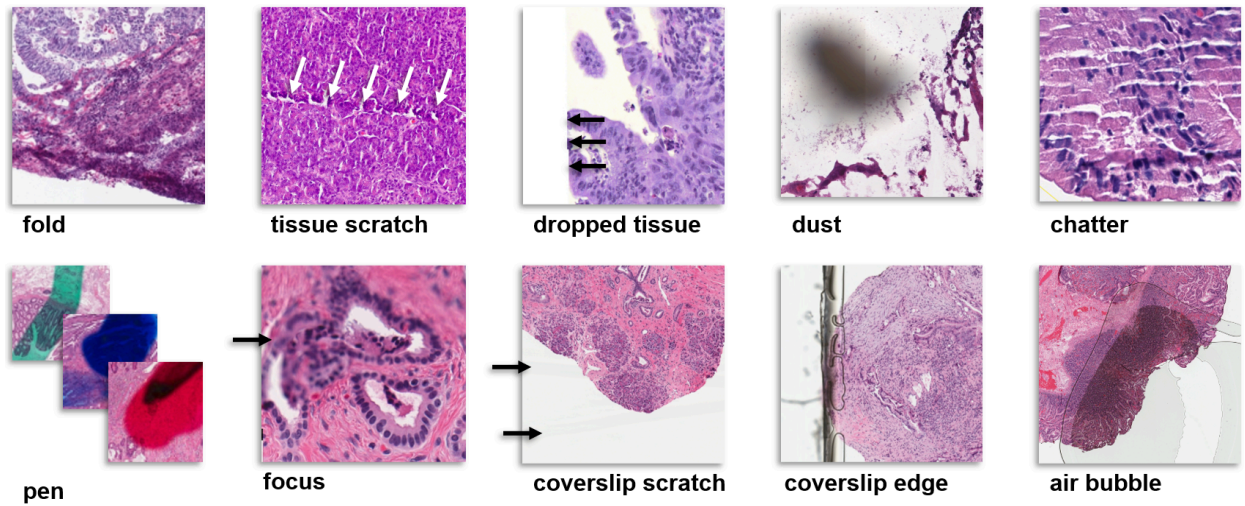

**Figure 2**. Representative examples of artifacts used for training the AI model.

| Tile Class (256px2) | 2.5x | 5x (L6) | 10x (L4) | 20x (L2) |
|---|---|---|---|---|
| tissue | 10246 | 11137 | 145893 | 177672 |
| focus | 996 | 1003 | 14140 | 14164 |
| pen | 704 | 769 | 6927 | 11204 |
| air bubble | 324 | 341 | 3644 | 4391 |
| dust | 430 | 545 | 3638 | 4211 |
| dropped tissue | 317 | 319 | 1460 | 2402 |
| fold | 268 | 357 | 1718 | 2051 |
| chatter | 99 | 107 | 1093 | 1128 |
| tissue scratch | 104 | 109 | 647 | 647 |
| debris | 19 | 19 | 147 | 147 |
| coverslip scratch | 18 | 18 | 94 | 94 |
| coverslip edge | 2 | 2 | 13 | 13 |
| other | 0 | 0 | 0 | 0 |
| Total Tiles | 13527 | 14726 | 179414 | 218124 |

**Table 1.** Count of tiles at varying magnification and static tilesize of 256 px² demonstrating decrease in tile count as magnification decreases and context increases.

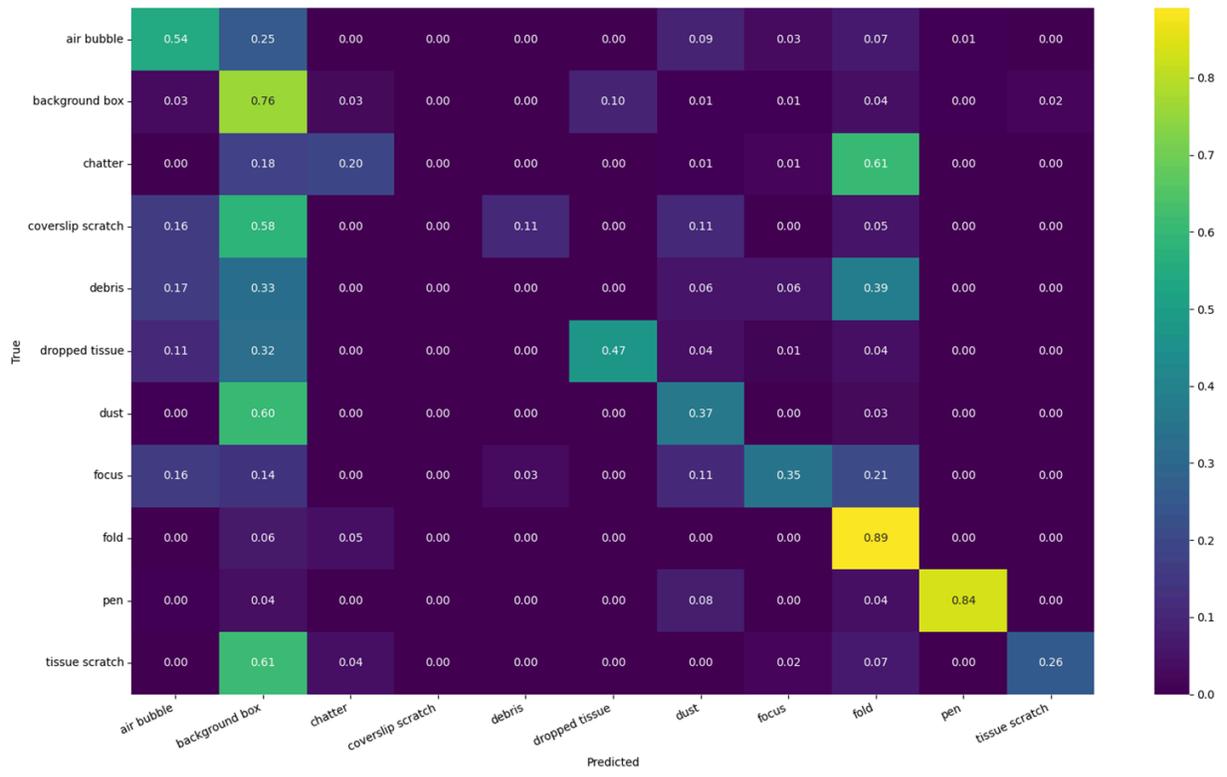

**Figure 3.** Confusion matrix shows the classification results of the best-performing configuration from the ablation studies (InceptionResNet, 256 px², L2). Performance is suboptimal on most artifacts, but shows promise on those with sufficient data. This configuration shows high true positive rate for *fold* (0.89) and *pen* (0.84) artifacts. Frequent misclassification of *chatter* and *debris* as *fold*, likely indicating data imbalance issues. Lower performance for *air bubble* (0.54), *dropped tissue* (0.47), *dust* (0.37), and *focus* (0.35), which also appear to suffer from data imbalance challenges.

| Model Architecture | Accuracy | Precision | Recall | F1 Score | Specificity |
|---|---|---|---|---|---|
| densenet | 0.402 | 0.314 | 0.376 | 0.252 | 0.938 |
| VGG16 | 0.836 | 0.288 | 0.332 | 0.285 | 0.972 |
| xception | 0.851 | 0.277 | 0.319 | 0.273 | 0.976 |
| inceptionV3 | 0.878 | 0.330 | 0.438 | 0.334 | 0.980 |
| inception_resnet | 0.879 | 0.330 | 0.406 | 0.330 | 0.968 |
| resnet50 | 0.888 | 0.354 | 0.402 | 0.368 | 0.977 |
| VGG19 | 0.888 | 0.317 | 0.392 | 0.328 | 0.981 |

**Table 2.** Performance metrics of individual models at L2-256 px². InceptionV3, InceptionResNet, ResNet50, and VGG19 showing roughly equivalent performance ahead of Xception, DenseNet, and VGG16. InceptionResNet was chosen due to its features addressing the vanishing gradient problem, complex feature extraction. We were confident that it is better equipped to learn the features in our data as the dataset increases in size and complexity.

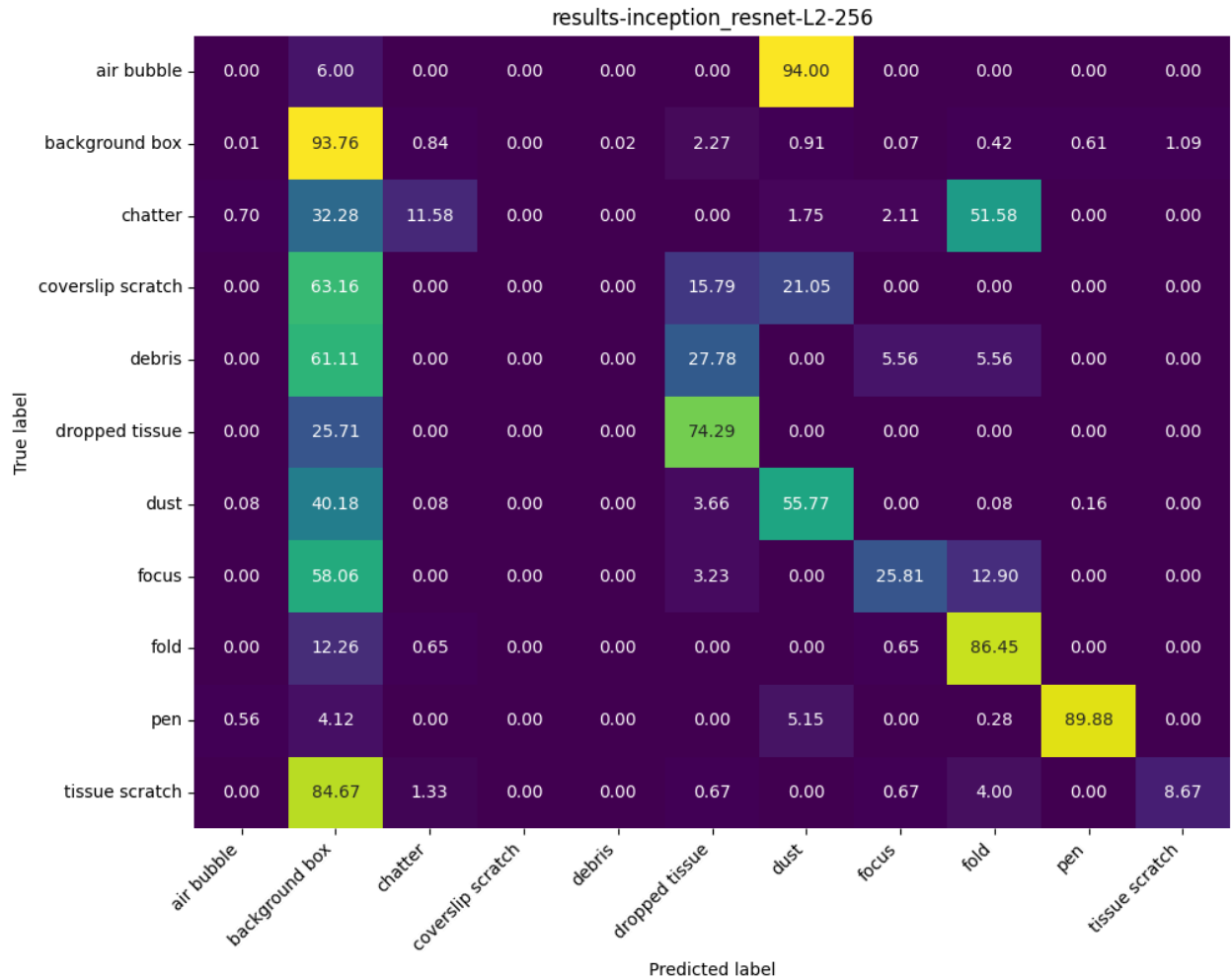

**Figure 4.** Adding green coloring to the background that was used to fill unscanned areas of slide improved the performance of *dropped tissue* at L2-256px$^2$ within the whole set. Data were augmented by direct copying tiles to meet the mean of that artifact at the slide level so that no one slide was overrepresented.

| Class | tilesize | Accuracy | Precision | Recall | F1 Score | Specificity | Total Count |
|---|---|---|---|---|---|---|---|
| air bubble | 128 | 0.97 | 0.44 | 0.99 | 0.61 | 0.97 | 1145 |
| chatter | 128 | 0.99 | 0.64 | 0.96 | 0.77 | 0.99 | 649 |
| coverslip scratch | 128 | 1.00 | 1.00 | 0.96 | 0.98 | 1.00 | 47 |
| debris | 128 | 1.00 | 1.00 | 0.93 | 0.97 | 1.00 | 184 |
| dropped tissue | 128 | 0.99 | 0.98 | 0.32 | 0.48 | 1.00 | 663 |
| dust | 128 | 0.93 | 0.17 | 0.98 | 0.30 | 0.93 | 811 |
| focus | 128 | 0.98 | 0.98 | 0.95 | 0.96 | 0.99 | 23393 |
| fold | 128 | 0.98 | 0.88 | 0.64 | 0.74 | 1.00 | 1807 |
| pen | 128 | 1.00 | 0.99 | 0.98 | 0.98 | 1.00 | 2329 |
| tissue scratch | 128 | 1.00 | 0.98 | 0.55 | 0.71 | 1.00 | 365 |
| air bubble | 256 | 1.00 | 1.00 | 1.00 | 1.00 | 1.00 | 300 |
| chatter | 256 | 1.00 | 0.99 | 1.00 | 1.00 | 1.00 | 225 |
| coverslip scratch | 256 | 1.00 | 1.00 | 1.00 | 1.00 | 1.00 | 20 |
| debris | 256 | 1.00 | 1.00 | 1.00 | 1.00 | 1.00 | 55 |
| dropped tissue | 256 | 0.98 | 0.43 | 1.00 | 0.61 | 0.98 | 233 |
| dust | 256 | 1.00 | 0.98 | 1.00 | 0.99 | 1.00 | 339 |
| focus | 256 | 0.98 | 1.00 | 0.93 | 0.96 | 1.00 | 5981 |
| fold | 256 | 1.00 | 0.96 | 1.00 | 0.98 | 1.00 | 656 |
| pen | 256 | 1.00 | 1.00 | 1.00 | 1.00 | 1.00 | 665 |
| tissue scratch | 256 | 1.00 | 0.99 | 1.00 | 1.00 | 1.00 | 161 |
| air bubble | 512 | 1.00 | 1.00 | 1.00 | 1.00 | 1.00 | 91 |
| chatter | 512 | 1.00 | 1.00 | 1.00 | 1.00 | 1.00 | 80 |
| coverslip scratch | 512 | 1.00 | 1.00 | 1.00 | 1.00 | 1.00 | 10 |
| debris | 512 | 1.00 | 1.00 | 1.00 | 1.00 | 1.00 | 19 |
| dropped tissue | 512 | 1.00 | 0.99 | 0.94 | 0.97 | 1.00 | 103 |
| dust | 512 | 1.00 | 1.00 | 1.00 | 1.00 | 1.00 | 149 |
| focus | 512 | 1.00 | 1.00 | 1.00 | 1.00 | 1.00 | 1556 |
| fold | 512 | 1.00 | 1.00 | 1.00 | 1.00 | 1.00 | 256 |
| pen | 512 | 1.00 | 1.00 | 1.00 | 1.00 | 1.00 | 202 |
| tissue scratch | 512 | 1.00 | 1.00 | 1.00 | 1.00 | 1.00 | 62 |

**Table 3.** Performance metrics at L2 magnification level for binary classification models for single artifacts.

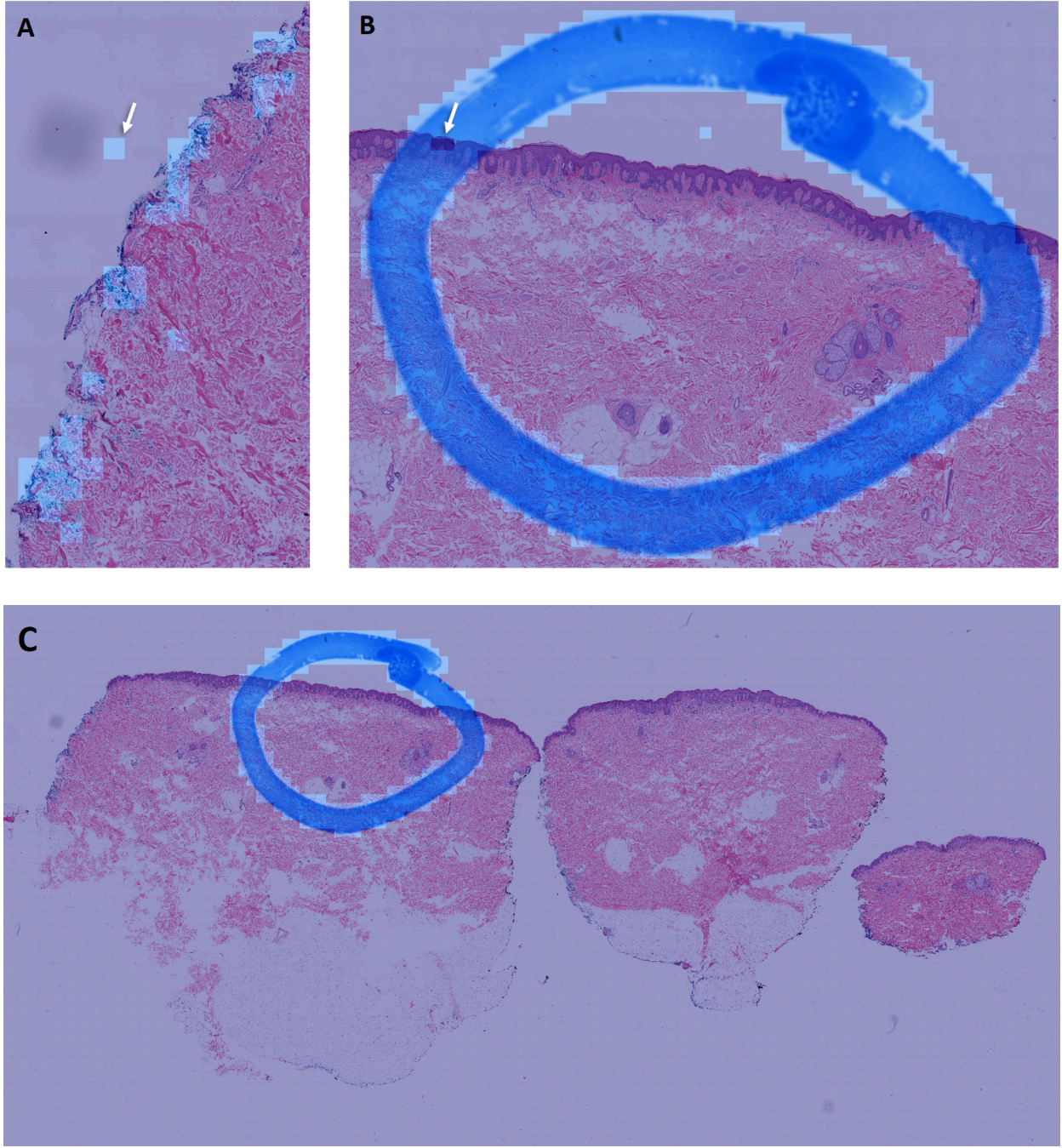

**Figure 5.** Images illustrating prediction results for the single artifact model trained on L2-128 px² to detect pen artifacts. Two representative tiles are highlighted to showcase challenges in artifact detection: **False Positive (A)**: A tile is incorrectly predicted as containing a pen artifact. The region in question contains blue tissue ink, which was applied to mark a resection margin. **False Negative (B)**: Two tiles tile that contain a darkly stained squamous epithelium layer

underlying a blue pen marking is incorrectly classified as negative (no pen artifact). The darkly stained epithelium in other cases annotated as negative may have contributed to the false negative at this magnification and tile size due to lack of context. **(C)** A model trained on L2-256 px² performed without these misclassifications.

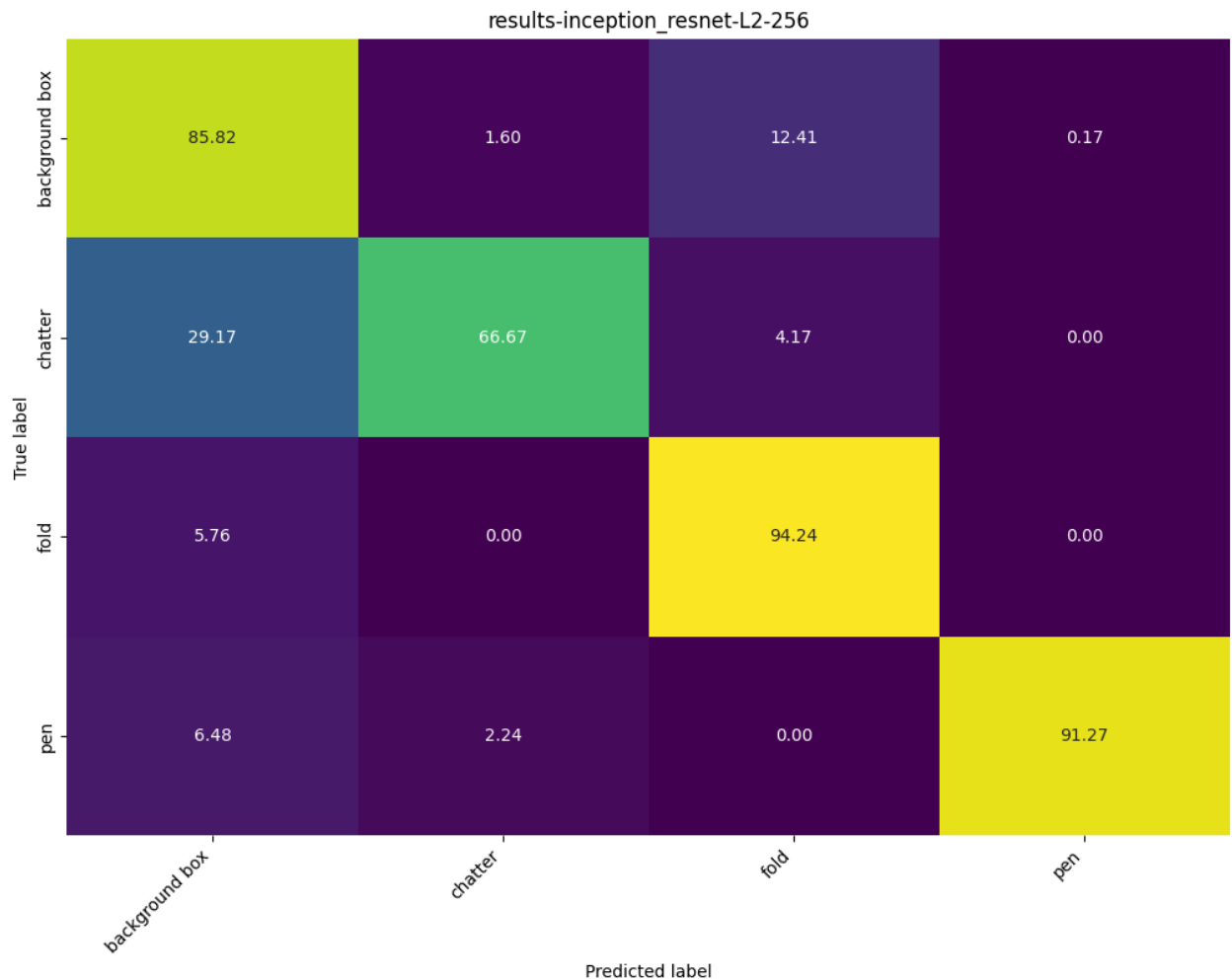

**Figure 6.** A limited model trained for *fold, pen,* and *chatter* artifacts. Confusion matrix showing excellent performance for fold and pen artifact detection at L2-256 px$^2$.

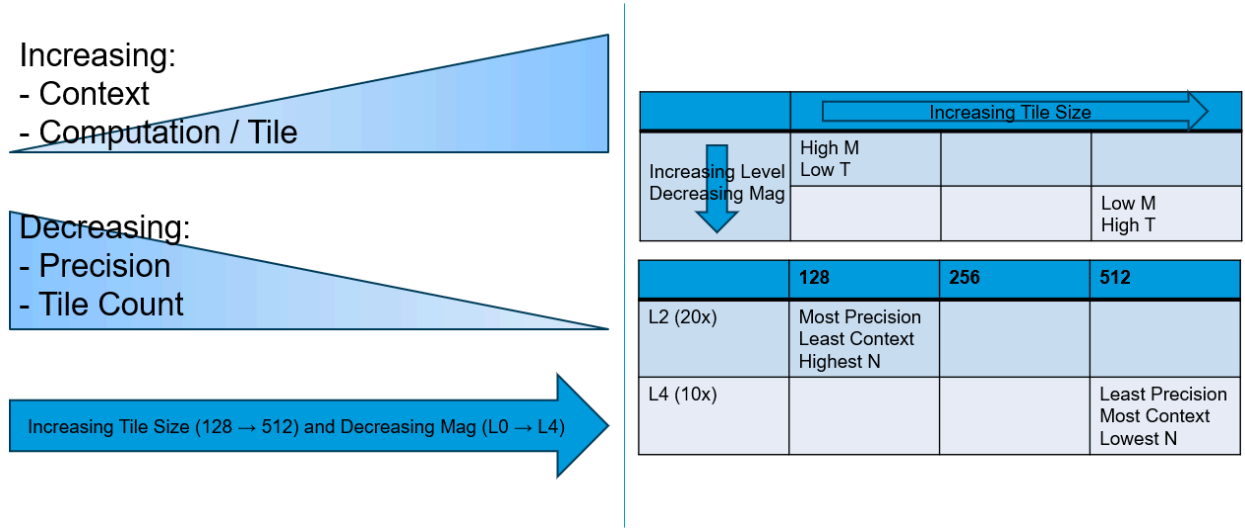

**Figure 7.** Increasing tilesize and decreasing magnification causes context in each tile to increase along with computation cost per tile. However, the total number of tiles also decreases along with precision. Lower magnification models were subject to variability due to low data volume *N*.

| Artifact vs. Background True Label | Prediction artifact | background box |
|---|---|---|
| air bubble | 66.67% | 33.33% |
| background box | 13.70% | 86.30% |
| chatter | 79.17% | 20.83% |
| coverslip scratch | 46.15% | 53.85% |
| dropped tissue | 100.00% | 0.00% |
| dust | 75.00% | 25.00% |
| focus | 58.06% | 41.94% |
| fold | 94.24% | 5.76% |
| pen | 97.01% | 2.99% |
| tissue scratch | 17.65% | 82.35% |

**Table 4.** True positive rate (Artifact Prediction) with acceptance of any artifact as a true positive. Data were augmented by direct copying tiles to meet the mean of that artifact at the slide level so that no one slide was overrepresented. If the top prediction was any artifact, it was counted

as "artifact" true positive for any other artifact. This demonstrates effective screening for chatter, dropped tissue, dust, fold, and pen.